\documentclass[aps,twocolumn,amsmath,amssymb]{revtex4-2}
\usepackage{graphicx} 
\usepackage{dcolumn} 
\usepackage{bm} 
\usepackage{braket} 
\usepackage[usenames]{color}
\usepackage{ulem}
\usepackage[
colorlinks=true, 
citecolor=blue, 
urlcolor=blue, 
linkcolor=blue,
setpagesize=false, 
bookmarks=false
]{hyperref}
\usepackage[english]{babel}

\begin{document}


\title{Photoinduced $\eta$-pairing correlation in the Hubbard ladder}
\author{Ryota Ueda, Kazuhiko Kuroki, and Tatsuya Kaneko}
\affiliation{Department of Physics, Osaka University, Toyonaka, Osaka 560-0043, Japan}
\date{\today}


\begin{abstract}
We investigate the staggered correlation of the on-site pairs, the so-called $\eta$-pairing correlation, induced by pump electric fields in the Hubbard model on the ladder lattice. 
Employing the time-evolution method based on exact diagonalization, we compute the photoinduced $\eta$-pairing correlation with different strengths of the interchain hopping. 
When the pump field is polarized along the chain direction, the $\eta$-pairing correlation is noticeably induced in the weakly coupled double-chain region, where the pairing property is similar to the $\eta$ pairing in the single-chain Hubbard model. 
On the other hand, the pump field polarized along the rung direction prominently induces the $\eta$-pairing correlation in the strong interchain coupling regime, where the pair correlations developing across rungs strongly contribute to the $\eta$ pairing. 
Based on the local rung approximation, which is valid in the strong interchain coupling regime, we discuss the origin of the photoinduced pair. 
\end{abstract}

\maketitle


\section{Introduction}
Light-matter couplings enable us to create and manipulate intriguing quantum phenomena in driven nonequilibrium systems~\cite{basov2017,giannetti2016,yonemitsu2008,ishihara2019,delatorre2021a}. 
For example, light-induced anomalous Hall effect has been reported in graphene~\cite{mclver2020}, and possible light-induced superconducting properties have been discussed in copper oxide and fullerene systems~\cite{fausti2011,mitrano2016}.
Floquet engineering, i.e., design of quantum states in periodically driven systems, is also a recent active research topic~\cite{oka2019,rudner2020,eckardt2017,bukov2015}.
In correlated many-body systems, pump electric fields can activate hidden degrees of freedom. 
In Mott-Hubbard systems, photoexcitation can induce the $\eta$-pairing states~\cite{kaneko2019}, which are the eigenstates of the Hubbard model possessing staggered correlations of the on-site pairs~\cite{yang1989,essler2005}. 
The light-induced enhancement of the $\eta$-pairing correlation has been demonstrated by the real-time simulations~\cite{kaneko2019,kaneko2020,ejima2020,ejima2022}, and the presence of the $\eta$-pairing phase has been verified in the photodoped Mott insulators~\cite{li2020,murakami2022a,murakami2023,murakami2023photoinduced,sarkar2023}. 
The nonequilibrium $\eta$ pairing has also been discussed in different types of driving schemes~\cite{kitamura2016,peronaci2020,tindall2020,tindall2021}.
These studies suggest that the $\eta$ degrees of freedom play a key role in the photoexcited and nonequilibrium states of the Hubbard model. 

Recently, there has been renewed interest in bilayer and ladder systems associated with the discovery of high-temperature superconductivity in pressurized La$_3$Ni$_2$O$_7$~\cite{sun2023,hou2023,zhang2023_arXiv2307.14819,sakakibara2023TE,wang2023}. 
In this material,
the distinctive electronic structures are made by strong interlayer bonding, which is a key factor for understanding the origin of superconductivity~\cite{luo2023,sakakibara2023,zhang2023_arXiv2307.15276,kaneko2023}.
Furthermore, the recent realization of the magnetically meditated hole pairing in the optical ladder lattice~\cite{hirthe2023} and possible strong pairing in doped mixed-dimensional Mott insulators~\cite{bohrdt2022} provide advanced insights into the physics of the bilayer/ladder systems. 
In these systems, quantum correlations across rungs involving anisotropic kinetic and magnetic properties between directions parallel and perpendicular to the chains (layers)~\cite{dagotto1992,troyer1996} enable the creation of entangled interchain (interlayer) states. 
In terms of light-matter coupling, since there are two possible polarization directions, parallel or perpendicular to the rung, external light may drive unique anisotropic properties in ladder and bilayer systems depending on the polarization direction. 
Given such background, the Hubbard ladder [a one-dimensional (1D) analog of the bilayer Hubbard model] is an attractive host to study light-induced phenomena, and it is natural to ask how $\eta$-pairing correlations appear in ladder systems.

In this paper, we investigate the $\eta$ pairing in the optically driven Hubbard ladder. 
By employing the time-evolution method based on exact diagonalization, we compute the $\eta$-pairing correlation with different ratios of the interchain hopping $t_{\perp}$ to the intrachain hopping $t_{\parallel}$. 
We find when the pump electric field is polarized along the chain direction, similarly to the 1D Hubbard chain, $\eta$ pairing is noticeably induced in the weakly coupled double-chain region (at $t_{\perp}/t_{\parallel} < 1$).  
On the other hand, the external field polarized along the rung direction induces the rung-like $\eta$ pairing in the strong interchain hopping regime (at $t_{\perp}/t_{\parallel} > 1$), where the pair correlations across rungs strongly contribute to the photoinduced pairing. 
Since the $\eta$ pairing with strong $t_{\perp}$ is unique to the ladder system, we discuss the origin of the photoinduced pair based on the local rung approximation. 

The rest of this paper is organized as follows. 
In Sec.~\ref{sec:model}, we introduce the Hubbard ladder model under the pump electric field and the numerical method we used.  
In Sec.~\ref{sec:results}, we show the calculated pair correlation functions, where the polarization and $t_{\perp}/t_{\parallel}$ dependences are presented. 
Then, we discuss the origin of the photoinduced pair using the local rung approximation. 
The summary is given in Sec.~\ref{sec:summary}.


\section{Model and Method} \label{sec:model}

The Hamiltonian of the Hubbard ladder is defined by

\begin{align}
    \begin{split}
        \hat{\mathcal{H}} =
        &- t_\parallel \sum_{j,\alpha,\sigma}(\hat{c}^\dagger_{j,\alpha;\sigma}\hat{c}_{j+1,\alpha;\sigma} + \rm H.c. ) \\
        &- t_\perp \sum_{j,\sigma}(\hat{c}^\dagger_{j,0;\sigma}\hat{c}_{j,1;\sigma} + {\rm H.c.})
        + U \sum_{j,\alpha} \hat{n}_{j,\alpha;\uparrow} \hat{n}_{j,\alpha;\downarrow} ,
    \end{split}
\end{align}
where $\hat{c}_{j,\alpha;\sigma}$ ($\hat{c}^\dagger_{j,\alpha;\sigma}$) is the annihilation (creation) operator for a fermion with spin $\sigma = \uparrow,\downarrow$ at site $j$ on chain $\alpha$ ($=0,1$), and $\hat{n}_{j,\alpha;\sigma} = \hat{c}^\dagger_{j,\alpha;\sigma} \hat{c}_{j,\alpha;\sigma}$.
$t_\parallel$ and $t_\perp$ are the hopping integrals along the chain and rung directions, respectively. 
$U$ $(> 0)$ is the on-site Coulomb repulsion. 
We consider a ladder of length $L$ at half-filling, where the number of particles $N$ is equal to the number of lattice sites $2L$.

Since the two-leg ladder is a bipartite lattice, we can introduce the $\eta$ operators~\cite{yang1989,essler2005}, which are given by
$\hat{\eta}^+_{j,\alpha} = (-1)^{j+\alpha} \hat{c}^\dagger_{j,\alpha;\downarrow} \hat{c}^\dagger_{j,\alpha;\uparrow}$, $\hat{\eta}^-_{j,\alpha} = (-1)^{j+\alpha} \hat{c}_{j,\alpha;\uparrow} \hat{c}_{j,\alpha;\downarrow}$, and $\hat{\eta}^z_{j,\alpha} = (\hat{n}_{j,\alpha;\uparrow} + \hat{n}_{j,\alpha;\downarrow} -1) / 2$.
These operators satisfy the SU(2) commutation relations $[\hat{\eta}^+_{j,\alpha} , \hat{\eta}^-_{j,\alpha}] = 2 \hat{\eta}^z_{j,\alpha}$ and $[\hat{\eta}^z_{j,\alpha} , \hat{\eta}^\pm_{j,\alpha}] = \pm \hat{\eta}^\pm_{j,\alpha}$, where $[A,B] = AB-BA$.
The total $\eta$ operators are defined by $\hat{\eta}^\pm = \sum_{j,\alpha} \hat{\eta}^\pm_{j,\alpha}$ and $\hat{\eta}_z = \sum_{j,\alpha} \hat{\eta}^z_{j,\alpha}$, which satisfy $[\hat{\mathcal{H}}, \hat{\eta}^+ \hat{\eta}^-] = [\hat{\mathcal{H}}, \hat{\eta}_z] = 0$.
These commutative properties indicate that the eigenstates of the Hubbard Hamiltonian are labeled by the quantum numbers of $\hat{\eta}^2$ [$=  (\hat{\eta}^+ \hat{\eta}^- + \hat{\eta}^- \hat{\eta}^+)/2 + \hat{\eta}_z^2$] and $\hat{\eta}_z$.
In this condition, the ground state of the Hubbard model at half filling ($\eta_z = 0$) is an eigenstate with $\eta=0$~\cite{kaneko2019}.

In order to induce $\eta$ pairing, we introduce a pump electric field via the Peierls substitution~\cite{peierls1933}.
When an external field is polarized along the chain direction, we introduce the vector potential $A_\parallel(t)$ by replacing
$\hat{c}^\dagger_{j,\alpha;\sigma}\hat{c}_{j+1,\alpha;\sigma} \rightarrow e^{-i q A_\parallel(t)} \hat{c}^\dagger_{j,\alpha;\sigma}\hat{c}_{j+1,\alpha;\sigma}$, where $q$ is the charge of a particle and the electric field is given by $E_\parallel(t)=-\partial_t A_\parallel(t)$.
The light velocity $c$, the Planck constant $\hbar$, and the lattice constant $a$ are set to 1, and we use $q=-1$ in our calculations.
Similarly, $\hat{c}^\dagger_{j,0;\sigma}\hat{c}_{j,1;\sigma} \rightarrow e^{-iqA_\perp(t)} \hat{c}^\dagger_{j,0;\sigma}\hat{c}_{j,1;\sigma}$ when the vector potential in the rung direction $A_\perp(t)$ is nonzero. 
In this paper, we consider a pump field
\begin{equation}
    \bm{A}(t) = \bm{A}_0 \exp \left[ - \frac{(t-t_0)^2}{2\sigma_p^2} \right] \cos[\omega_p (t-t_0)] ,
\end{equation}
where $\bm{A}(t) = A_\parallel(t) \bm{e}_\parallel + A_\perp(t) \bm{e}_\perp$.
$A_0 = |\bm{A}_0|$ is the amplitude of the vector potential, $\omega_p$ is the frequency, and time $t_0$ and $\sigma_p$ correspond to the center and width of the time-dependent pulse, respectively.
The external field $\bm{A}(t) \ne 0$ transiently breaks the commutation relation between the operator $\hat{\eta}^2$ and the time-dependent Hamiltonian $\hat{\mathcal{H}}(t)$, which allows the photoinduced $\eta$ pairing~\cite{kaneko2019}.  

The wave function $\ket{\Psi(t)}$ under the pump electric field is numerically obtained by solving the time-dependent Schr\"{o}dinger equation $i \frac{d}{dt} \ket{\Psi(t)} = \hat{\mathcal{H}}(t) \ket{\Psi(t)}$. 
We employ the exact diagonalization (ED) method for the initial ground state $\ket{\psi_0}$.
The sequential time evolution $\ket{\Psi(t + \delta t)} \simeq e^{-i \hat{\mathcal{H}}(t) \delta t}\ket{\Psi(t)}$ with small time step $\delta t$ is carried out by  the Lanczos technique~\cite{park1986,lu2012,hashimoto2016}, where the exponential is expanded as $e^{-i \hat{\mathcal{H}} \delta t}  \simeq \sum_{m=0}^{M_{\rm L}-1} [ (-i\delta t)^m /m!] ( \hat{ \mathcal{H}} )^m$. 
$M_{\rm L}$ corresponds to the number of the iterations in the Lanczos algorithm, and a large enough $M_{\rm L}$ (with small $\delta t$) provides a numerically precise result. 
In our calculations, we set $t_\parallel$ ($t^{-1}_\parallel$) as a unit of energy (time). 
The number of the Lanczos vectors for time evolution is $M_{\rm L} =15$
and we adopt $\delta t =0.01/t_{\parallel}$. 
We use the $6\times 2$ site cluster (i.e., $L=6$) with periodic boundary conditions along the chain direction.
We set $t_0=10/t_\parallel$ and $\sigma_p=2/t_\parallel$ in the pump field at all calculations.

We numerically observe the $\eta$-pairing properties using the pair correlation functions 
\begin{align}
     & P(x,0;t) = \frac{1}{2L} \sum_{j,\alpha} \Braket{\Psi(t) | \left( \hat{\Delta}^\dagger_{j+x,\alpha} \hat{\Delta}_{j,\alpha} + {\rm H.c.}  \right) | \Psi(t)},
    \\
     & P(x,1;t) = \frac{1}{2L} \sum_{j,\alpha} \Braket{\Psi(t) | \left( \hat{\Delta}^\dagger_{j+x,\bar{\alpha}} \hat{\Delta}_{j,\alpha} + {\rm H.c.}  \right) | \Psi(t)},
\end{align}
where $\hat{\Delta}_{j,\alpha} = \hat{c}_{j,\alpha;\uparrow} \hat{c}_{j,\alpha;\downarrow}$ is the operator of the on-site pair and $\bar{\alpha}$ denotes the opposite chain index to $\alpha$ (i.e., $\bar{0} = 1$ and $\bar{1} = 0$).
$P(x,0;t)$ and $P(x,1;t)$ represent intra- and interchain correlations of the on-site pairs, respectively.
These correlation functions can be denoted as $P(x,y;t)$ or $P(\bm{r};t)$ with $\bm{r} = (x,y)$. 
The Fourier transform of the pair correlation function may be given by 
\begin{equation}
    P(q_\parallel,q_\perp;t) = \sum_{x,y} e^{i(q_\parallel x + q_\perp y)} P(x,y;t).
\end{equation}
Note that $q_\perp = 0$ or $\pi$ corresponds to the parity along the rung direction~\cite{endres1996}. 
In this reciprocal space notation, the staggered correlation $P(q_\parallel \!=\! \pi,q_\perp \!=\! \pi;t) = \sum_{x,y} (-1)^{x + y} P(x,y;t)$, which is proportional to $\langle \Psi(t) | \hat{\eta}^2 | \Psi(t) \rangle$ at half filling ($\eta_z=0$), corresponds to the $\eta$-pairing correlation.
To decompose the $\eta$-pairing correlation into intra- and interchain contributions, we define
\begin{align}
     & P_{{\rm intra}}(t) = \sum_{x>0} (-1)^x P(x,0;t),
    \\
     & P_{\rm inter}(t) = \sum_{x} (-1)^{x+1} P(x,1;t), 
\end{align}
respectively.   
Since the on-site part $P(x=0,y=0;t)$ corresponds to twice the double occupancy $n_d(t) = (1/2L) \sum_{j,\alpha} \braket{ \Psi(t) | \hat{n}_{j,\alpha;\uparrow} \hat{n}_{j,\alpha;\downarrow} | \Psi(t) }$, we find
\begin{equation}
    P(q_\parallel \!=\! \pi,q_\perp \!=\! \pi;t) = 2n_d (t)  + P_{\rm intra}(t) + P_{\rm inter}(t).
    \label{eq:Ppi_decompose}
\end{equation}


\section{Results} \label{sec:results}

\subsection{Photoinduced $\eta$ pairing}

\begin{figure}[t]
    \begin{center}       
        \includegraphics[width=\columnwidth]{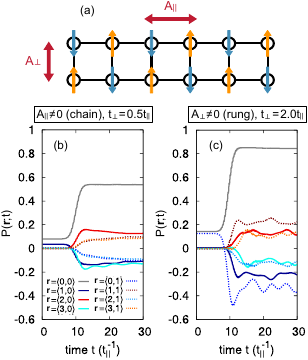}        
        \caption{(a) Schematic figure of the Hubbard ladder and polarization directions of the pump field.
         (b) Time evolution of the pair correlation function $P(\bm{r};t)$ when $A_{\parallel}(t) \ne 0$  [and $A_{\perp}(t)=0$] at $U=8t_\parallel$ and $t_\perp=0.5t_\parallel$, where $A_0=0.3$ and $\omega_p=7.2t_\parallel$ are used in $\bm{A}(t)$. 
         (c) Time evolution of the pair correlation functions $P(\bm{r};t)$ when  $A_{\perp}(t) \ne 0$  [and $A_{\parallel}(t)=0$] at $U=8t_\parallel$ and $t_\perp=2.0t_\parallel$, where $A_0=0.2$ and $\omega_p=9.8t_\parallel$ are used in $\bm{A}(t)$. 
         The solid (dotted) line represents the intrachain (interchain) component of the pair correlation function.
         }
        \label{fig1}
    \end{center}
\end{figure}

\begin{figure}[t]
    \begin{center}
        \includegraphics[width=\columnwidth]{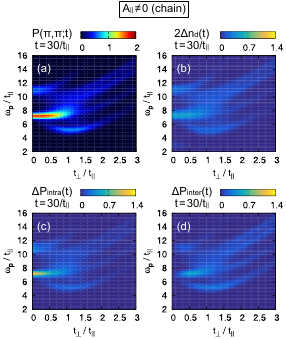}        
        \caption{$t_\perp$ and $\omega_p$ dependence of (a) $P(\pi,\pi;t)$, (b) $2 \Delta n_d(t)$, (c)$\Delta P_{\rm intra}(t)$, and (d)$\Delta P_{\rm inter}(t)$ at $t = 30/t_\parallel$ when the pump field is polarized along the chain direction [$A_{\parallel}(t) \ne 0$  and $A_{\perp}(t)=0$], where $U=8t_\parallel$ and $A_0=0.3$ are used.
        }
        \label{fig2}
    \end{center}
\end{figure}

\begin{figure}[t]
    \begin{center}      
        \includegraphics[width=\columnwidth]{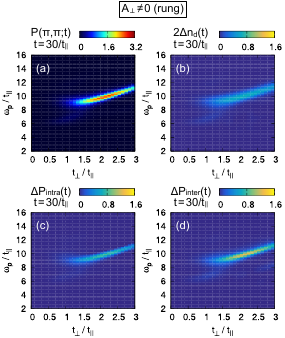}        
        \caption{$t_\perp$ and $\omega_p$ dependence of (a) $P(\pi,\pi;t)$, (b) $2 \Delta n_d(t)$, (c)$\Delta P_{\rm intra}(t)$, and (d)$\Delta P_{\rm inter}(t)$ at $t = 30/t_\parallel$ when the pump field is polarized along the rung direction [$A_{\parallel}(t) = 0$  and $A_{\perp}(t)\ne 0$], where $U=8t_\parallel$ and $A_0=0.2$ are used.   
        }
        \label{fig3}
    \end{center}
\end{figure}

Here, we show the calculated pair correlation function $P(\bm{r};t)$ when the pump field is polarized along the chain or rung direction [see Fig.~\ref{fig1}(a)]. 
We use the optimal parameter sets of $t_{\perp}$ and $\bm{A}(t)$ for the $\eta$ pairing in Fig.~\ref{fig1}, and the parameter dependence of the pair correlation is shown in Figs.~\ref{fig2} and \ref{fig3}.  

In Fig.~\ref{fig1}(b), we plot the time evolution of $P(\bm{r};t)$ when the weakly coupled chains at $t_{\perp}$=0.5$t_\parallel$ are driven by the external field polarized along the chain direction [i.e., $A_{\parallel}(t) \ne 0$  and $A_{\perp}(t)=0$]. 
Similarly to the 1D single chain~\cite{kaneko2019}, we find a strong enhancement of $P(\bm{r}=(0, 0); t)$ corresponding to the enhancement of the double occupancy $n_d(t)$. 
In addition to this local component, $P(\bm{r}\ne(0, 0);t)$ is also enhanced by $A_{\parallel}(t)$. 
$P(\bm{r};t)$ at $x + y =$ odd is negative while $P(\bm{r};t)$ at $x + y =$ even is positive, indicating the staggered pair correlation of the $\eta$ pairing~\cite{kaneko2019}. 
In the ladder system, we find that the development of the intrachain component $P(x,0;t)$ is faster than that of the interchain component $P(x,1;t)$. 
This is because the external field $A_{\parallel}(t)$ preferentially creates the doublon (doubly occupied site)  and holon (empty site) in the same chain, and then these carriers move across rungs by the interchain hopping $t_{\perp}$. 
Hence, the primal driving force of the $\eta$-pair in the weakly coupled chains is essentially the same as that in the 1D single chain. 

Figure~\ref{fig1}(c) shows the time evolution of $P(\bm{r},t)$ when the strongly coupled chains at $t_{\perp}$=2.0$t_\parallel$ are driven by the external field polarized along the rung direction [i.e., $A_{\parallel}(t) = 0$  and $A_{\perp}(t) \ne 0$]. 
Similarly to Fig.~\ref{fig1}(b), we find the signatures of the photoinduced $\eta$ pairing. 
However, there are several differences. 
In comparison with Fig.~\ref{fig1}(b), the enhancement of the double occupancy $n_d (t)$ is larger. 
This indicates that the open boundary or nature of the two-site Hubbard model in the rung direction is favorable for the doublon creation. 
As for the time profile, the interchain component at $\bm{r}=(0,1)$ is firstly enhanced, and then the intrachain and nonlocal ($ x > 1 $) correlations are developed. 
In Fig.~\ref{fig1}(c), the interchain components at $\bm{r} = (0,1)$ is larger than the intrachain pair correlation. 
Since the magnitudes of the intrachain components are comparable to these values in Fig.~\ref{fig1}(b), the interchain components strongly induced by $A_{\perp}(t)$ reinforce the $\eta$-pairing correlation. 
In contrast to the previous case, the doublon and holon are created by $A_{\perp}(t)$ at the same rung, and then the carrier motion along the chain direction via $t_{\parallel}$ may promote the nonlocal part of the correlation. 

In Figs.~\ref{fig2} and \ref{fig3}, we plot the $\eta$-pairing correlation $P(q_{\parallel}= \pi,q_{\perp}= \pi;t)$ and its components at $t=30/t_{\parallel}$ (after pulse irradiation) in the plane of $t_{\perp}$ and $\omega_p$. 
Figure~\ref{fig2}(a) shows the $\eta$-pairing correlation when the external field is polarized along the chain direction [i.e., $A_{\parallel}(t) \ne 0$  and $A_{\perp}(t)=0$]. 
Note that $P(\pi,\pi;t=0)=0$ since the ground state is the eigenstate with the quantum number $\eta=0$ and $\eta_z=0$. 
As shown in Fig.~\ref{fig2}(a), the $\eta$-pairing correlation is prominently induced by $A_{\parallel}(t)$ around $\omega_p / t_{\parallel} = 7$ in the small $t_{\perp}$ region while the induced pair correlation decreases as $t_{\perp}$ is increased.  
This tendency implies that the $\eta$-pairing property in the 1D single chain mainly causes the $\eta$ pairing in the ladder. 
In order to identify the main contributor to the $\eta$ pairing in the ladder, we decompose $P(\pi,\pi;t)$ into the double occupancy $n_d(t)$, intrachain component $P_{\rm intra}(t)$, and interchain component $P_{\rm inter}(t)$ [see Eq.~(\ref{eq:Ppi_decompose})]. 
In Figs.~\ref{fig2}(b)-\ref{fig2}(d), we plot $\Delta n_d(t)$, $\Delta P_{\rm intra}(t)$, and $\Delta P_{\rm inter}(t)$ at $t=30/t_{\parallel}$, where $\Delta X(t) = X(t)-X(0)$ indicates the difference of the quantity $X$ from its equilibrium value at $t=0$. 
We find that $n_d$ is enhanced at the same spot as the $\eta$-pairing correlation is induced.  
In addition to $n_d(t)$, the intrachain component $P_{\rm intra}(t)$ in the small $t_{\perp}$ regime is prominently activated by $A_{\parallel}(t)$.  
This means that the strongly induced $P(\pi,\pi;t)$ at $t_{\perp}/t_{\parallel} < 1$ is mainly caused by nonlocal intrachain correlations in $P_{\rm intra}(t)$.  
This chain-like $\eta$-pairing correlation becomes smaller at $t_\perp / t_\parallel > 1$, indicating that the formation of the strong rung bond is unfavorable for the $\eta$ pairing induced by $A_{\parallel}(t)$.  

Figure~\ref{fig3} shows the $t_{\perp}$ and $\omega_p$ dependence of the correlation functions when the external field is polarized along the rung direction [i.e., $A_{\parallel}(t) = 0$  and $A_{\perp}(t) \ne 0$].  
In contrast to the chain-like $\eta$ pairing shown in Fig.~\ref{fig2}(a), the enhancement of $P(\pi,\pi;t)$ is not noticeable in the small $t_{\perp}$ regime. 
On the other hand, $\eta$-pairing correlation is significantly light-induced at $t_{\perp}/t_{\parallel}>1$. 
The maximum value of $P(\pi,\pi;t)$ around $t_{\perp}/t_{\parallel}\sim 2$ in Fig.~\ref{fig3}(a) is nearly $1.5$ times as large as the maximum value observed in Fig.~\ref{fig2}(a) [see the maximum of the color bars, which are 2 in Fig.~\ref{fig2}(a) and 3.2 in Fig.~\ref{fig3}(a)].  
As shown in Figs.~\ref{fig3}(b)-\ref{fig3}(d), the interchain component $P_{\rm inter}(t)$ at $t_{\perp}/t_{\parallel}>1$ mainly contributes to the enhancement of the $\eta$-pairing correlation.  
This rung-like $\eta$ pairing should be contrasted with the chain-like $\eta$ pairing shown in Fig.~\ref{fig2}. 
As seen in Fig.~\ref{fig1}(c), $A_{\perp}(t)$ activates the strong interchain component $P(x,1;t)$ in addition to the intrachain component $P(x,0;t)$. 
Because of the additional $P(x,1;t)$ contribution, the $\eta$-pairing correlation driven by $A_{\perp}(t)$ is enhanced relative to the correlation of the chain-like $\eta$ pairing induced by $A_{\parallel}(t)$. 
These results suggest that the formation of the rung bond is favorable for $\eta$ pairing when the external field is polarized along the rung direction.

\subsection{Local rung approximation}

\begin{figure}[t]
    \begin{center}        
        \includegraphics[width=\columnwidth]{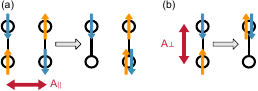}                
        \caption{Schematic figures of the local rung approximation for the doublon-holon creations induced by the external fields applied along the (a) chain and (b) rung directions.
        }
        \label{fig4}
    \end{center}
\end{figure}

Here, we discuss the origin of the photoinduced pair in the large $t_{\perp} / t_{\parallel}$ regime since the chain-like $\eta$ pairing at $t_{\perp} / t_{\parallel} \ll 1$ can be understood in terms of the photoinduced $\eta$ pairing in the single Hubbard chain, whose mechanism has been revealed in Ref.~\cite{kaneko2019}. 
When $t_\perp/t_\parallel \gg 1$, the Hubbard ladder with $L \times 2$ sites can be approximated as the $L$ independent two-site Hubbard models~\cite{endres1996}. 
In this local rung approximation (LRA), the essential physical properties of the ladder can be estimated by the eigenstates in the two-site Hubbard model. 
Figure~\ref{fig4} schematically shows two patterns of light-induced doublon-holon creation in the LRA.
When the external field is applied along the chain direction as shown in Fig.~\ref{fig4}(a), one of two particles in the spin-singlet bond transfers to the adjacent rung (by the perturbative weak $t_{\parallel}$), and one doublon and one holon are created along the chain direction.
On the other hand, when the external field is applied along the rung direction as shown in Fig.~\ref{fig4}(b), one of two particles moves to the rung direction, and one doublon and one holon are created within the single rung.  
Here, we consider the optical excitation energies for these doublon-created states and compare them with the results obtained by the ED-based calculations. 

\begin{figure}[t]
    \begin{center}  
        \includegraphics[width=\columnwidth]{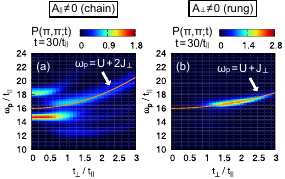}          
        \caption{$t_\perp$ and $\omega_p$ dependence of $P(\pi,\pi;t \!=\! 30/t_\parallel)$ at $U=16t_\parallel$.
        (a) $P(\pi,\pi;t)$ after the pulse $A_{\parallel}(t) \ne 0$  and $A_{\perp}(t)=0$ with $A_0=0.3$ and (b) $P(\pi,\pi;t)$ after the pulse $A_{\parallel}(t) = 0$  and $A_{\perp}(t) \ne 0$ with $A_0=0.2$. 
         The orange solid lines indicate $\omega_p = U+2J_{\perp}$ in (a) and $\omega_p = U+J_{\perp}$ in (b), where $J_{\perp} = 4t_\perp^2 / U$. 
         }
        \label{fig5}
    \end{center}
\end{figure}

First, we consider the case in Fig.~\ref{fig4}(a).  
The ground state of two particles in the two-site Hubbard model is the spin-singlet state whose eigenenergy is $-J_{\perp} = - 4t_{\perp}^2/U$ in the strong coupling limit ($U \gg t_{\perp}$). 
Hence, the energy of two spin-singlet rungs in equilibrium (before the pump) is $E^{\rm (a)}_{\rm g} \simeq -2J_{\perp}$. 
When the pump field $A_{\parallel}(t)$ is applied, one of the two rungs possesses one particle while the other rung possesses three particles [see Fig.~\ref{fig4}(a)]. 
The energies of the one-particle states are $E_1 = \mp t_{\perp}$ while the energies of the three-particle states are $E_3 = U \pm t_{\perp}$, where upper and lower signs correspond to even and odd parities along the rung direction, respectively. 
Because the external field polarized along the chain direction does not change the parity along the rung direction, the parities of the one-particle and three-particle states must be the same.
Hence, the excited energy of the two rungs shown in Fig.~\ref{fig4}(a) is given by $E^{\rm (a)}_{\rm e} = \mp t_\perp + (U \pm t_\perp) = U$ in total. 
According to these energies, the optical gap of the case in Fig.~\ref{fig4}(a) is given by $\Delta^{\rm (a)} = E^{\rm (a)}_{\rm e} - E^{\rm (a)}_{\rm g} \simeq U + 2J_{\perp}$. 
This energy gap suggests that the $\eta$-pairing correlation can be activated at $\omega_p = \Delta^{\rm (a)} \simeq U + 2J_{\perp}$. 
Let us compare the energy estimated by the LRA with the actual numerical data. 
Figure~\ref{fig5}(a) shows $P(\pi,\pi;t)$ at $t=30/t_{\parallel}$ in the $t_{\perp}$ and $\omega_p$ plane, where we set $U=16t_\parallel$ because our estimation of the optical gap is accurate when $U \gg t_{\perp}$. 
In the region at $t_{\perp}/t_{\parallel} > 1$, we indeed find that $U + 2J_\perp$ shows good agreement with $\omega_p$ at which the $\eta$-pairing correlation is generated. 
Because the LRA is not valid when $t_{\perp}/t_{\parallel} \ll 1$, the peak positions of $P(\pi,\pi;t)$ around $t_{\perp}/t_{\parallel} = 0$ deviate from $U + 2J_\perp$.   
However, even around  $t_\perp / t_\parallel = 1$, the hot spot of $P(\pi,\pi;t)$ and the curve $\omega_p = U + 2J_\perp$ almost coincide. 
In reality, $t_\parallel$ is not zero, and complicated multiple factors can be involved in the optical transition. 
The appearance of weak multiple bands at $t_{\perp}/ t_{\parallel} > 1$ besides the $U + 2J_\perp$ line in Fig.~\ref{fig5}(a) may be caused by these minor excitation entities. 

Next, we consider the case in Fig.~\ref{fig4}(b).  
Before the pump, the ground-state energy of one spin-singlet rung is $E^{\rm (b)}_{\rm g} \simeq -J_{\perp}$. 
The eigenenergy of the optically allowed one doublon and one holon state (i.e., $\eta$-pair state) in the two-site Hubbard model is $E^{\rm (b)}_{\rm e} =U$. 
Hence, the optical gap of the case in Fig.~\ref{fig4}(b) is given by $\Delta^{\rm (b)} = E^{\rm (b)}_{\rm e} - E^{\rm (b)}_{\rm g} \simeq U + J_{\perp}$, suggesting that the $\eta$-pairing correlation can be activated at $\omega_p = \Delta^{\rm (b)} \simeq U + J_{\perp}$. 
Figure~\ref{fig5}(b) shows the comparison between the LRA and the numerical data. 
As expected, the peak positions of the induced $P(\pi,\pi;t)$ at $t=30/t_{\perp}$ show good agreement with the line $\omega_p = U + J_{\perp}$ estimated by the LRA.  
These agreements indicate that the LRA at $t_{\perp} / t_{\parallel} > 1$ is a very valid approach for identifying the optimal $\omega_p$ of the photoinduced pair in the Hubbard ladder. 

Finally, we comment on the polarization dependence of the strength of the $\eta$-pairing correlation at $t_{\perp}/t_{\parallel} > 1$.  
While the intensity regions of $P(\pi,\pi;t)$ at $t_{\perp}/t_{\parallel} > 1$ show good agreement with the lines estimated by the LRA, the $\eta$-pairing correlation induced by $A_{\parallel}(t) \ne 0$ in Fig.~\ref{fig5}(a) is not so large in comparison with the pair correlation induced by $A_{\perp}(t) \ne 0$ in Fig.~\ref{fig5}(b). 
This is because $t_{\parallel}$ is small relative to $t_{\perp}$ in this region, and the doublon generation along the chain direction meditated by $t_{\parallel}$ is inefficiently induced by $A_{\parallel}(t)$.  
On the other hand, since the doublon generation along the rung direction meditated by $t_{\perp}$ can be strongly induced by $A_{\perp}(t)$, in Fig.~\ref{fig5}(b), we observe the prominently induced $\eta$-pairing correlation at $t_{\perp}/t_{\parallel} > 1$. 
In Fig.~\ref{fig5}(b), the $\eta$-pairing correlation is significantly enhanced at $1.5<t_{\perp}/t_{\parallel} < 2.0$. 
In the independent rung limit ($t_{\parallel} / t_{\perp} \rightarrow 0$), the $\eta$-pairing correlation must be suppressed because a weak $t_{\parallel}$ relative to $t_{\perp}$ is unfavorable for the spatial extension of the correlation along the chain direction, where the nonlocal ($x > 1$) pair correlations cannot contribute to $P(\pi,\pi;t)$. 
For this reason, the $\eta$-pairing correlation in Fig.~\ref{fig5}(b) can be enhanced in the intermediate $t_{\perp}/t_{\parallel}$ regime.   
Hence, we can interpret the intensity of $P(\pi,\pi;t)$ by considering the contributions of $t_{\parallel}$ in the LRA. 


\section{Summary} \label{sec:summary}

We have investigated the $\eta$ pairing in the optically driven Hubbard ladder using the time-evolution method based on ED.
As in the 1D Hubbard chain, we have observed the light-induced enhancement of the $\eta$-pairing correlation in the Hubbard ladder when the pump field is polarized along the chain direction.  
Moreover, we have shown that the pair correlation induced by the pump field polarized along the rung direction is larger than the correlation induced by the field polarized along the chain direction. 
This consequence is mainly caused by the strong interchain component of the pair correlations. 
This rung-like $\eta$ pairing observed in the large $t_\perp / t_\parallel$ region is unique in the ladder system.
Finally, the origin of the photoinduced pair in the strong $t_{\perp}$ regime has been clarified using the LRA.

The pump electric field can also enhance charge correlations associated with the doublon creation. 
In the 1D photodoped Mott insulator, the $\eta$-pairing correlation is dominant when the nearest-neighbor Coulomb interaction $V$ is weak whereas the charge-density-wave correlation becomes dominant when $V$ is large~\cite{murakami2022a,murakami2023}.   
As in the 1D chain, the intersite interaction $V$ may enhance the charge-density-wave correlation in the photodoped ladder system. 
Experimentally, $\eta$ pairing has not been observed up to date. 
Since the maximum value of the induced $\eta$-pairing correlation in the Hubbard ladder is larger than that of the single chain limit at $t_{\perp}=0$, the ladder lattice can be a promising host of the photoinduced $\eta$ pairing. 
For instance, the ladder-type cuprates~\cite{sparta2006} can be candidates for the host of the rung-like $\eta$ pairing. 
If we can set up a similar situation in a many-body simulator such as an optical lattice, we may approach the $\eta$ pairing using the ladder structure. 
These are open issues for the future. 


\begin{acknowledgments}
We thank D. Gole\v{z}, Z. Lenar\v{c}i\v{c}, and  M. Sarkar for fruitful discussions.
This work was supported by Grants-in-Aid for Scientific Research from JSPS, KAKENHI Grants No. JP18K13509, No. JP20H01849, and No. JP22K04907.
R.U. was supported by Program for Leading Graduate Schools: ``Interactive Materials Science Cadet Program''.  
\end{acknowledgments}

\begin{figure}[b]
    \begin{center}  
        \includegraphics[width=\columnwidth]{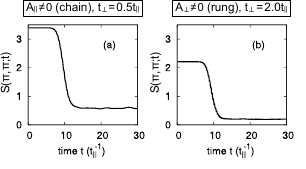}          
        \caption{Time evolution of the antiferromagnetic correlation function $S(\pi,\pi;t)$ when (a) $A_{\parallel}(t) \ne 0$  and (b) $A_{\perp}(t) \ne 0$.  
        The parameters used in (a) and (b) are the same as the parameters in Figs.~\ref{fig1}(b) and \ref{fig1}(c), respectively. 
        }
        \label{fig6}
    \end{center}
\end{figure}


\appendix

\section{Spin Correlation}
In this Appendix, we show the antiferromagnetic (AF) correlation function
\begin{align}
S(q_\parallel \!=\! \pi,q_\perp \!=\! \pi;t) = \sum_{x,y} (-1)^{x + y} S(x,y;t)
\end{align}
given by
\begin{align}
    & S(x,0;t) = \frac{1}{2L} \sum_{j,\alpha} \Braket{\Psi(t) | \hat{\sigma}^z_{j+x,\alpha} \hat{\sigma}^z_{j,\alpha}  | \Psi(t)},
   \\
    & S(x,1;t) = \frac{1}{2L} \sum_{j,\alpha} \Braket{\Psi(t) | \hat{\sigma}^z_{j+x,\bar{\alpha}} \hat{\sigma}^z_{j,\alpha} | \Psi(t)},
\end{align}
where $\hat{\sigma}^z_{j,\alpha}=\hat{n}_{j,\alpha;\uparrow} - \hat{n}_{j,\alpha;\downarrow}$. 
Figure~\ref{fig6} plots the time evolution of $S(q_\parallel \!=\! \pi,q_\perp \!=\! \pi;t)$, where the parameters used in 
Figs.~\ref{fig6}(a) and \ref{fig6}(b) are the same as Fig.~\ref{fig1}(b) and \ref{fig1}(c), respectively.
Figures~\ref{fig6}(a) and \ref{fig6}(b) are the results when the pump electric fields are polarized along the chain and rung directions, respectively. 
The initial states (at $t=0$) possess strong AF correlations. 
Since the formation of the rung spin-singlet suppresses the AF correlation, the initial state at $t_{\perp}/t_{\parallel}=2$ [Fig.~\ref{fig6}(b)] has a smaller AF correlation than that at  $t_{\perp}/t_{\parallel}=0.5$ [Fig.~\ref{fig6}(a)]. 
In both cases, the pump electric fields reduce the AF correlations. 
This suppression is mainly caused by the photoinduced doublon creation. 
Because the double occupancy is strongly enhanced by $A_{\perp}(t)$ as shown in Fig.~\ref{fig1}(c), the AF correlation after the pulse irradiation in Fig.~\ref{fig6}(b) is smaller than that in Fig.~\ref{fig6}(a).


\bibliography{Reference}

\end{document}